\documentclass[prl,amsmath,amssymb,showpacs,twocolumn,floats,superscriptaddress]{revtex4-1}
\usepackage{graphicx,morefloats,slashed}
\usepackage{subfigure}
\usepackage{bm}
\usepackage{color}

\begin{document}
\title{Boundaries Determine the Formation Energies of Lattice Defects in Two-Dimensional Buckled Materials}
\author{Sandeep K. Jain}
\email{S.K.Jain@uu.nl}
\affiliation
{Institute for Theoretical Physics, Universiteit Utrecht, Leuvenlaan 4, 3584 CE Utrecht, The Netherlands}
\author{Vladimir  Juri\v ci\'c}
\email{juricic@nordita.org}
\affiliation
{Nordita, Center for Quantum Materials, KTH Royal Institute of Technology,
and Stockholm University, Roslagstullsbacken 23, S-106 91 Stockholm, Sweden}
\affiliation{Institute for Theoretical Physics, Universiteit Utrecht, Leuvenlaan 4, 3584 CE Utrecht, The Netherlands}
\author{Gerard T. Barkema}
\affiliation
{Institute for Theoretical Physics, Universiteit Utrecht, Leuvenlaan 4, 3584 CE Utrecht, The Netherlands}
\affiliation
{Instituut-Lorentz, Universiteit Leiden, Niels Bohrweg 2, 2333 CA Leiden, The Netherlands}

\begin{abstract}
{Lattice defects are inevitably present in two-dimensional materials, with direct implications on their physical and chemical properties. We show that the formation energy of a lattice defect in buckled two-dimensional crystals is not uniquely defined as it takes different values for different boundary conditions even in the thermodynamic limit, as opposed to their perfectly planar counterparts. Also, the approach to the thermodynamic limit follows a different scaling: inversely proportional to the logaritm of the system
size for buckled materials, rather than the usual power-law approach. In graphene samples of $\sim 1000$ atoms, different boundary conditions can cause differences exceeding 10 eV.  Besides presenting numerical evidence in simulations, we show that the universal features in this behavior can be understood with simple bead-spring models. Fundamentally, our findings imply that it is necessary to specify the boundary conditions for the energy of the lattice defects in the buckled two-dimensional crystals to be uniquely defined, and this may explain the lack of agreement in the reported values of formation energies in graphene. We argue that boundary conditions may also have impact on other physical observables
such as the melting temperature.
}
\end{abstract}

\maketitle

Lattice irregularities in the form of defects, such as dislocations and grain boundaries, are quite generically present in crystalline lattices. Usually, defects have a direct impact on the various properties of the material; for instance, in graphene they reduce the mobility \cite{Chen2009}, change Young's modulus \cite{Zandiatashbar2014,Lapez-2014} and the fracture behavior
\cite{Grantab2010}. A fundamental property characterizing a lattice defect is its formation energy, with the crucial importance for their behaviour, e.g. the defects' migration and healing \cite{Skowron2015}. On the other hand, two-dimensional crystals have a natural tendency to buckle out of the crystalline plane to relieve the stress \cite{Fasolino2007,banhart-review-2011,Zhang2014}. For perfectly confined two-dimensional materials, the formation energy of a lattice defect does not depend on the boundary conditions, but only on the type of the defect, and in that sense is uniquely defined. However, the question arises whether this fundamentally important feature of the lattice defects changes in buckled crystals, and in particular whether the boundaries affect the defects' energy.

In this Letter, we show that this is indeed the case by studying the formation energy of the defects in both simple, analytically tractable buckled one- and two-dimensional bead-spring models, as well as in numerical simulations of graphene, a paradigmatic representative of a two-dimensional buckled crystal. In particular, we find that unlike two-, and three- dimensional materials where the formation energy of a lattice defect, such as an SW defect, is well defined, in buckled two-dimensional materials different boundary conditions give rise to different values of the formation energy of the defect in the thermodynamic limit. Moreover, while the finite-size correction in the energy scales as inversely proportional to the system size for one-, two-, and three-dimensional materials, we show that this scaling for buckled sheet-type materials is given by the inverse of logarithm of the system size.

\begin{figure*}
\begin{center}
\includegraphics[width=\textwidth, height=12cm]{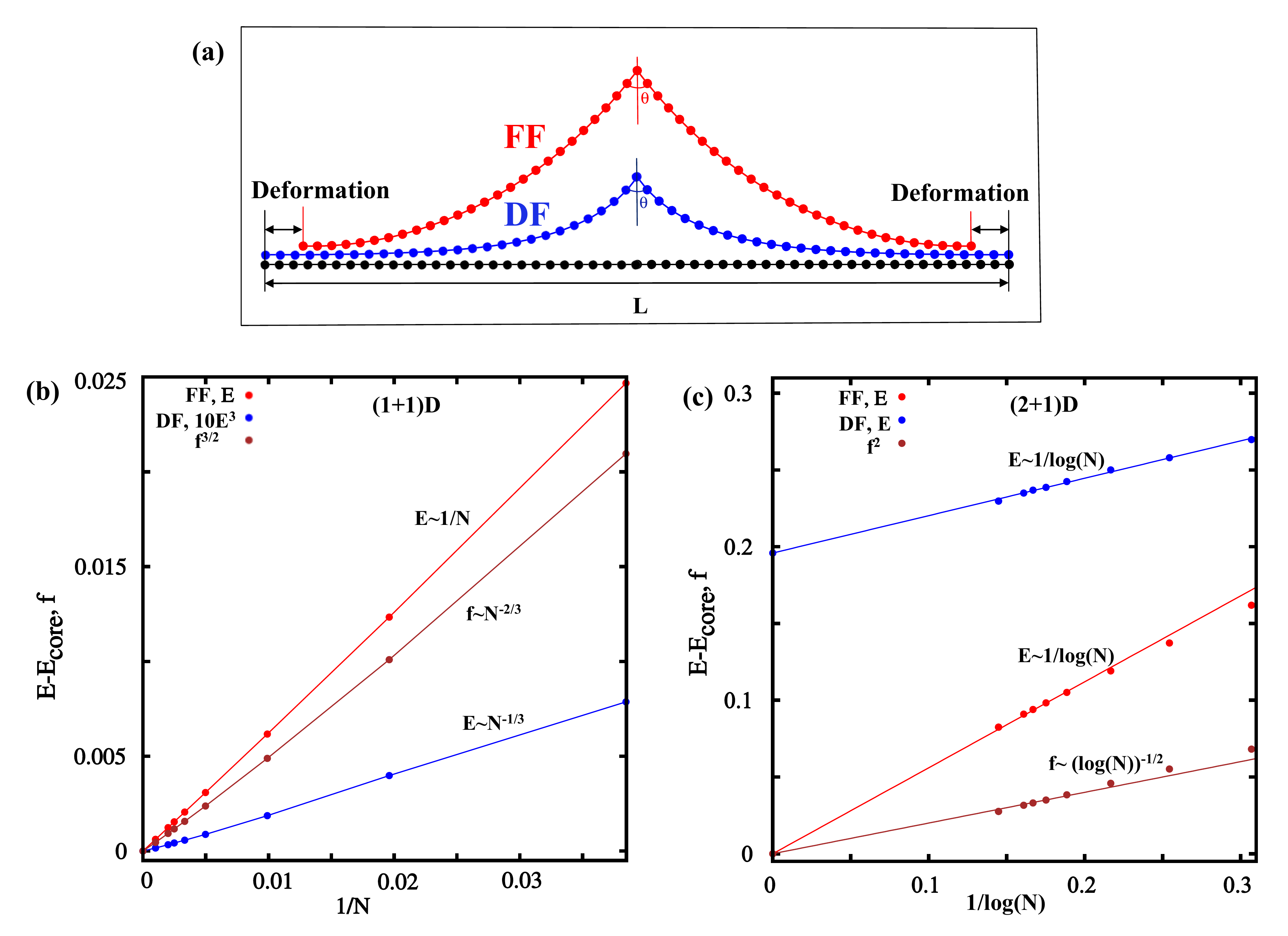}
\caption{ Illustration of force-free (FF) and deformation-free (DF) boundaries and calculated defect energy as a function of the system size for both (1+1)D and (2+1)D models. (a) Sketch of the elastic string model that accounts for the boundary effects on the formation energy of the defects. In case of DF boundaries, the introduction of the defect does not change the total length $(L)$ since a force is acting on the boundaries to keep the sample at constant density. In the case of FF boundaries, the density of the sample does change through the change of the length. (b) Finite-size correction to the energy and
force for both the FF and DF boundaries in (1+1)D model. The numerical data points are fitted well by the analytically predicted scaling of the energy for DF (blue line) and FF (red line) boundaries and force (brown line). (c) Finite-size correction to the energy and force for both the FF and DF boundaries in (2+1)D model. The numerical data points are fitted well by the analytically predicted scaling of
the energy for DF (blue line) and FF (red line) boundaries and square of the force (brown line). Here, we use the values of the parameters in the Hamiltonian (\ref{eq:hamiltonian}) $\lambda=\kappa=1$.}
\label{figure_1}
\end{center}
\end{figure*}

To describe this boundary effect on the formation energy of the defects in buckled crystals, we first consider a simple model of a string of $N$ atoms with length $L=N$, connected with elastic springs
and a defect created at the center of the string by making the bond angle with the $y$-axis equal to $\theta$ $\neq 0$, Fig.\ 1(a). This string is embedded in two-dimensional space and in this way we allow for the buckling in the model. In this (1+1)D model, the energy of the defect configuration is minimized for two most commonly used boundary conditions: force-free (FF) boundaries which relax the global planar stress, and deformation-free (DF) boundaries  which fix the density of the atoms to the crystalline density, Fig.\ 1(a).  We use the Hamiltonian
\begin{align}\label{eq:hamiltonian}
H=E_{core}+\lambda\sum_{i}(r_{i}-1)^2+\kappa\sum_{i}(\phi_{i+1}-\phi_{i})^2\nonumber\\
  -f\left(\sum_{i}r_{i}\cos(\phi_{i})-L\right),
\end{align}
where $r_i$ is the bond length between two neighbouring atoms $i-1$ and
$i$, and $\phi_i$ is the angle of this bond with respect to the $x$-axis.
For simplicity, we set the core energy of defect $E_{core}=0$.  The elastic constants
in the Hamiltonian are defined as: $\lambda$ is the bond stretching
constant, $\kappa$ is the bond bending constant, and $f$ is the force
acting on the boundaries.  At the FF boundary condition the energy is
minimized for $r_i=1$ and $\phi_{i}=\phi_{0}(-1+\frac{i}{N})$,
which leads to the finite-size energy scaling of $\sim 1/N$. In Fig.\
1(b) numerical values of FF energy calculations are shown (points) and
are in a very good agreement with the analytical solution (fitted with
line). Furthermore, DF boundary conditions yield a minimum energy for
$r_{i}=1+{f}/{(2\lambda)}$ and $\phi_{i}=\phi_{0} \exp{(-\alpha i)}$
with $\alpha=\sqrt{fr_{i}/2\kappa}$. These solutions in turn yield
forces with finite-size scaling of the form $f\sim N^{-2/3}$ while the
energy scales as $E\sim N^{-1/3}$. We have also performed the numerical
simulations for DF boundary conditions, and these results  are in
agreement with the analytical ones (Fig.\ 1(b)). More importantly, this
very simple model already yields a different scaling of the energy with
the system size for different boundaries, a feature also prominent in the (2+1)D model,
which we consider next.

To obtain the defect formation energy and its dependence on the system
size in two-dimensional space, we extended the one-dimensional model in
two dimensions in a rotationally-symmetric manner. We analytically solve the (2+1)D
model, as shown in the Supplementary Online Material (SOM) \cite{SOM}, and find that for FF boundary conditions the
energy scales as $\sim {1 /\log(N)}$ with system size. Furthermore,
at DF boundary conditions, the force scales as $f\sim {1/\sqrt{\log(N)}}$
whereas the energy scales as $E\sim {1/\log(N)}$ with a constant offset,
which is the formation energy of the defect in the thermodynamic limit. In
Fig.\ 1(c), we show the numerical calculations of both the energy and force
within this simple model. The data points are fitted with analytical
predictions, and show very good agreement. The most striking result here
is that both boundary conditions yield finite-size corrections of the form
${1/\log(N)}$, on top of a constant offset. In the limit of infinite
system size, FF and DF boundaries therefore yield different formation
energies.  This very simple model captures an essential feature of the
formation energy of a lattice defect in a buckled two-dimensional crystal,
which is its dependence
on the boundary conditions. Furthermore, the same model also produces the
finite-size scaling of the energy as found in our computer simulations
on graphene, which we present next.

To further demonstrate this effect,
we numerically study the formation energy of a single Stone-Wales (SW) defect, made of
a pair of pentagon-heptagon rings obtained
when four hexagons are transformed by a bond transposition of
$90^\circ$, in a graphene sheet buckled in the out-of-plane direction,
as shown in Fig.\ 2. We consider FF and DF boundary conditions, both of which
are periodic as commonly used in simulations.
Our results show that with DF boundaries the formation energy for the SW defects
is always significantly higher than with FF boundaries, and such boundaries
therefore strongly favor defect-free configurations of buckled graphene
samples. Contrary to the natural intuition, the energy difference
persists in the thermodynamic (infinite size) limit, as shown in Fig.\ 3, even though
all individual atomic positions become indistinguishable between the
two types of boundaries.
In finite-size samples, this gap is more pronounced,
as is especially the case for separated dislocations with FF versus DF boundary conditions (Fig.\ S3 in SOM) \cite{SOM}, where it can exceed $10$ eV.
Finite-size effects remain even in very large samples,
since the finite-size corrections in the energy decrease inversely proportional to the logarithm of the system size.
In contrast, if all atoms are confined to a purely two-dimensional plane, both FF and DF boundaries quickly converge to the same formation energy which is much higher than in the buckled samples. Finite-size corrections in this case decrease much faster, inversely proportional to the system size. Apparently,
buckling introduces strong finite-size effects, with boundary effects
that do not vanish in the thermodynamic limit.
Our results therefore imply that both the formation energy of the lattice defects
and its dependence on the size of the buckled graphene samples are not
well defined without specifying the boundary conditions, counter to the
conventional wisdom \cite{Skowron2015}.

We simulated structures of a graphene membrane with an SW defect.
Structural relaxation and energy computation are based
on a recently developed semi-empirical elastic potential for graphene
\cite{Sandeep2015}.  Eight different geometries were used in our
simulations.  They differ in the orientation of the SW defect relative
to the boundaries ($0{^\circ}$ and $60^{\circ}$), the buckling modes
(sine and cosine), and the types of boundaries (DF and FF).  The two
inequivalent initial bonds give rise to the two SW defects oriented
by $60^\circ$ relative to each other (Fig. 2(a) and 2(b)). The system is
then relaxed and to relieve the stress it buckles perpendicularly to
the flat graphene plane with the two possible buckling configurations,
sine and cosine, Fig.\ 2(c) and 2(d), while the density of carbon atoms is kept fixed (DF)
and relaxed (FF) [only FF shown in Fig.\ 2].

\begin{figure}[htb]
\includegraphics[width=8cm, height=6.2cm]{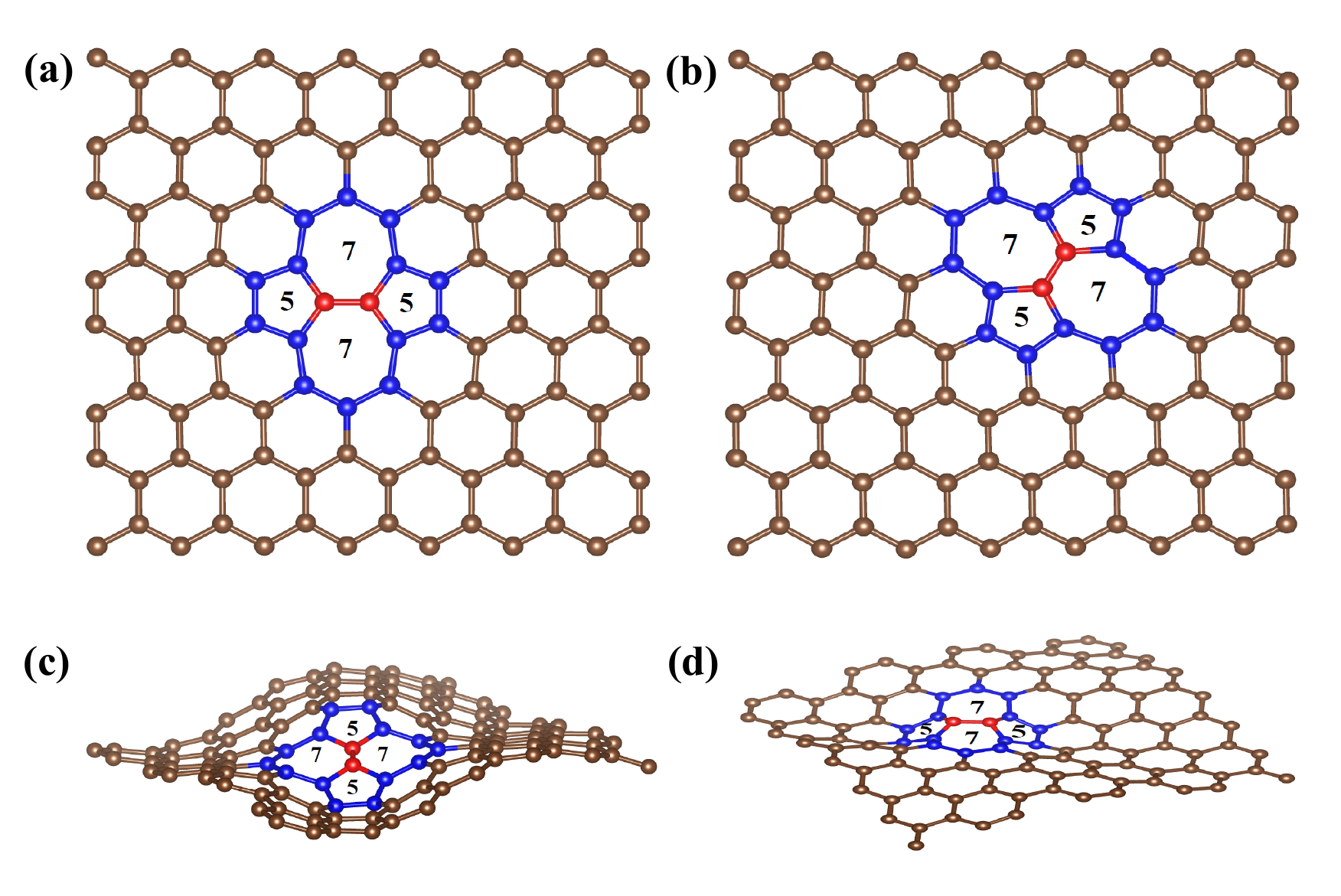}
\caption{Structure of the graphene sample with a single SW
defect.  The two different orientations of the defect are shown: (a) 0$^\circ$ and (b) 60$^\circ$.
Two different buckling modes represent (c) sine-type buckling and (d) cosine-type
buckling. The two configurations are shown from different viewing angles.  }
\label{figure_2}
\end{figure}

The calculated formation energies of a single SW defect in a buckled
graphene sheet for different system sizes are shown in Fig.\  3.
Its scaling with the system size is given by
\begin{equation}
E_{{\rm SW}}(N)=E_0+F(N),
\end{equation}
where $E_0$ is the energy contribution of the defect in an infinite
(square) system, and $F(N)$ describes finite size corrections, with
lateral sample size $L$ and a number of carbon atoms $N\sim L^2$.  For the
computational methods, see SOM \cite{SOM}. We first
observe for the eight structures that extrapolation to the infinite
system size produces four different values for the formation energy $E_0$
of the defect; the dependence on the orientation of the defect vanishes,
in agreement with the intuitive expectation based on the equivalence of
the $sp^2$ carbon bonds. On the other hand, the defect energy depends
on both the buckling configuration, and most notably, on the type of the
boundary of the sample. In particular, the DF boundaries, in which the density
of the carbon atoms is fixed to the crystalline value, always give a higher formation
energy of the defect than FF boundaries, see Table 1. Therefore, boundary conditions
play a crucial role in determining the formation energy of the defects.

\begin{table}[h]
\footnotesize
\begin{center}
\begin{tabular}{c|c|c}
\hline
\hline
{Defect and boundary type} & {$E_{0}$ (eV)}& {F(N)}\\
\hline
{FF, sin SW, 60 deg} & 2.87 & ${1}/{\sqrt{N}}$ *\\
\hline
{DF, sin SW, 60 deg} & 3.05 & ${1}/{\log(N)}$ \\
\hline
{FF, sin SW, 0 deg} & 2.87 & ${1}/{\sqrt{N}}$ *\\
\hline
{DF, sin SW, 0 deg} & 3.05 & ${1}/{\log(N)}$ \\
\hline
{FF, cos SW, 60 deg} & 3.02 & ${1}/{\sqrt{N}}$ * \\
\hline
{DF, cos SW, 60 deg} & 3.15 & ${1}/{\log(N)}$ \\
\hline
{FF, cos SW, 0 deg} & 3.02 & ${1}/{\sqrt{N}}$ * \\
\hline
{DF, cos SW, 0 deg} & 3.15 & ${1}/{\log(N)}$ \\
\hline
\hline
\end{tabular}
\caption{Formation energy $E_0$ of an SW defect in graphene in the thermodynamic limit,
and form of the corresponding leading finite-size corrections for different orientations, (0$^\circ$ or 60$^\circ$ with respect to the periodic directions, see Fig. 2(a),(b)), different buckling
(sine or cosine, see Fig. 2(c),(d)), and different boundaries (FF or DF).
Note that $E_0$ does not depend on defect orientation, but
does depend on the buckling mode as well as on the type of the boundary
conditions.  The leading finite-size correction in the formation energy
scales as ${1}/{\log(N)}$, with varying amplitude.
The lowest formation energy ($2.87$ eV) is for the configuration with
FF boundaries and sine-type buckling, whereas the highest ($3.15$ eV)
is for the DF boundaries with cosine-type buckling. Most importantly,
the formation energies in the thermodynamic limit for DF and FF
boundaries differ by 0.18 eV for sine-type buckling.\\
* In the case of FF boundaries, finite-size corrections for sample sizes
studied here (up to 137616 atoms) are dominated by the scaling factor of
${1}/{\sqrt{N}}$ \cite{Sandeep2015}, but a correction $\sim {1}/{\log(N)}$
with a small prefactor cannot be excluded. }
\label{2dTable}
\end{center}
\end{table}

This effect is especially pronounced when taking into account the finite
size of the graphene samples. As shown in Fig.\ 3(a), there is a notable
difference in the formation energy of the SW defects of up to 30\% between
the samples with DF and FF boundaries at the size of $N\sim 10^4$ atoms.
More importantly, the finite-size correction to the defect energy, $F(N)$,
scales as $1/\log N$ for DF boundaries. Therefore, DF boundaries besides
giving higher formation energy of the defects in the thermodynamic limit,
also give rise to its slow decrease with the system size. On the other hand, as shown in Fig. 3(b),
when the buckling is completely suppressed, the energy of the defect
in the thermodynamic limit converges to a common value independently
of the type of boundaries, with finite size correction $F(N)= C/N$
in which the prefactor $C$ differs for both types of boundaries and
the defect orientations. Notice also that in the flat graphene sheet,
the DF boundaries give the largest energy for the defect formation in
finite-size samples.

\begin{figure}[htb]
\includegraphics[width=8cm, height=11cm]{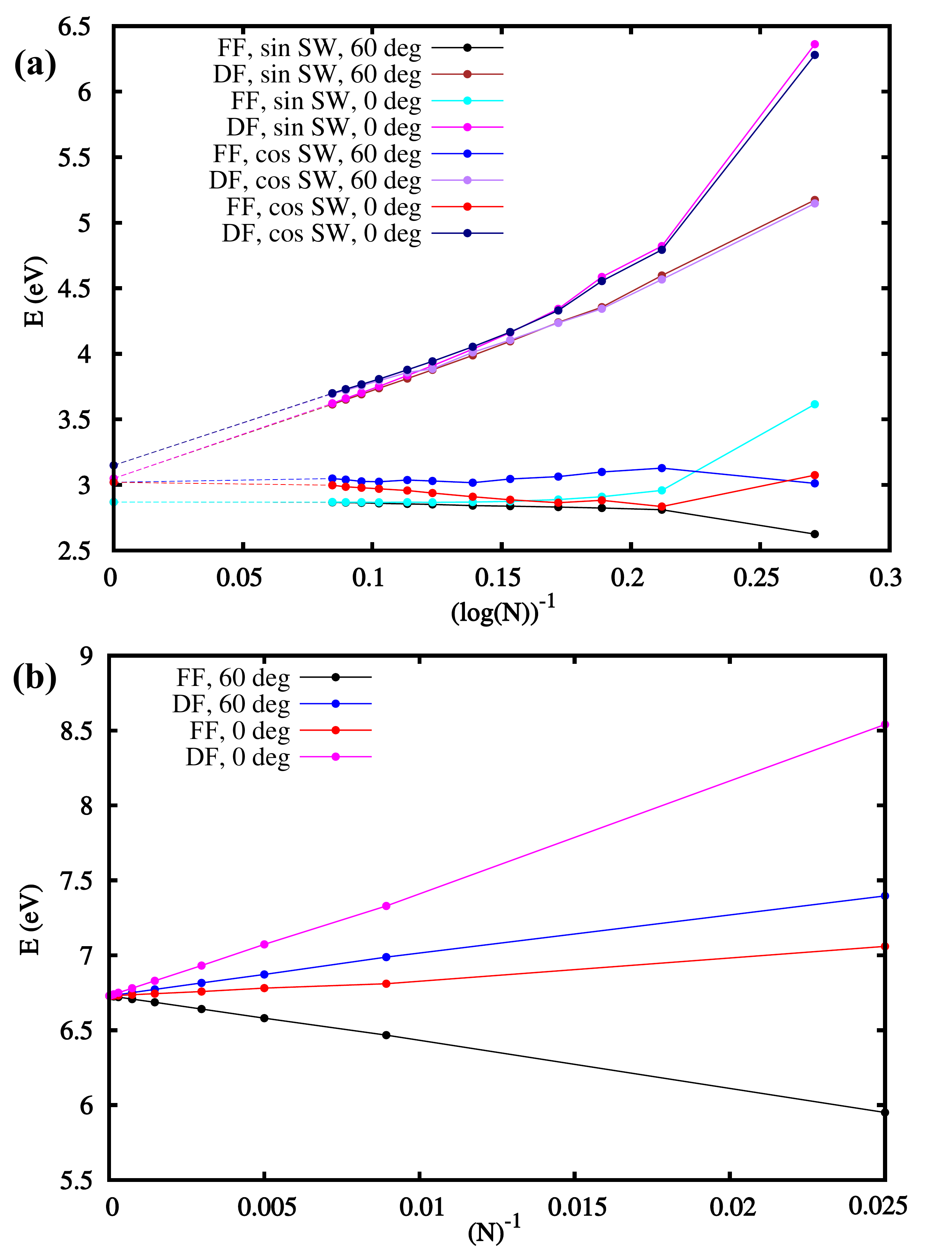}
\caption{ Formation energy as a function of the graphene sample size
with a single SW defect for both buckled (sine and cosine) and flat configurations,
with different boundaries (DF and FF) and defect orientations ($0^\circ$ and $60^\circ$).
(a) In buckled graphene, the formation
energy of the SW defect converges to four different values, determined by
the boundary condition and buckling mode; different orientations do not
influence the formation energy in the thermodynamic limit. Finite-size
corrections scale as $1/\log(N)$, with different prefactors for
different boundaries, buckling modes and defect orientations. (b)
In flat graphene, the formation energy of the SW defect converges to the same value ($6.73$
eV) irrespective of the orientation or the boundary condition, with finite-size
corrections scaling as $1/N$. }
\label{figure_3}
\end{figure}

The effects of the boundaries are even more pronounced when considering the energy of a dislocation pair as a function of the distance, see SOM \cite{SOM}. The size of the energy difference between the FF and DF boundaries for a dislocation pair can be of the order of 10 eV (Fig.\ S3). Moreover, the form of the potential between the dislocations depends heavily on the type of boundaries, implying a strong dependence of the melting temperature of graphene \cite{Los2015,Zakharchenko2011} on the boundary conditions. For FF boundary conditions the energy of a dislocation pair as a function of separation in a buckled graphene membrane quickly becomes constant as predicted by Seung and Nelson in the inextensional limit \cite{Seung1988}. The strain field around the core of a dislocation becomes short-ranged when buckling is allowed and therefore the energy converges to a finite value. On the other hand, for DF boundary conditions the energy of a dislocation pair in the buckled crystal increases with separation and this behaviour is consistent with the results obtained from a different elastic potential \cite{Los2015,Carlsson2011}. The increase in the energy in this case is lower than logarithmic, as predicted by Seung and Nelson. The strain field around the core of a dislocation does not become localized in the case of DF boundaries since a constant stretching force is applied at the boundaries in order to keep the atom density fixed and this could be the origin of the boundary effects. Furthermore, the force at the boundaries decreases with increasing system size, but at the same time the length of the boundary increases, and the combined effect on the defect's energy apparently is a constant offset as shown by our (2+1)D analytical model and numerical simulations on graphene.

Another qualitative way to understand our results, which at first glance seem surprising, is that a defect such as SW, locally deforms the membrane thereby reducing the "footprint" in the 2D plane. With FF boundary conditions, the system can simply shrink to the reduced footprint, but with DF boundary conditions it cannot, resulting in significant stress. The latter raises the energy, even in the thermodynamic limit.

Our work demonstrates the crucial importance of boundaries for
determining the formation energy of the lattice defects.  Boundary effects may also be partly responsible for the large variation of the reported formation energies of defects in numerical
simulations on the graphene lattice \cite{Skowron2015,Ertekin2009}.  Simple models for
an elastic string and a membrane embedded in a higher-dimensional space
suggest their independence of the lattice geometry and the model,
and in that sense they may represent a universal feature of the low-dimensional buckled crystals.
Our findings may be relevant to graphene samples where SW defects \cite{Hashimoto2004,Meyer2008,Kotakoski2011} and  grain boundaries
 have been observed \cite{Huang2011,Yazyev2014,Tsen2013}.
Finally, our study opens up a route to
investigate this boundary effect on the defects' energy in other
two-dimensional crystalline materials,  such as Mo$_2$C \cite {Xu2015}, as well as in recently synthesized silicene \cite{Vogt2012,Feng2012}, germanene \cite{Bianco2013}, and stanene \cite{Zhu2015}.

We acknowledge the financial support by the FOM-SHELL-CSER program (12CSER049). This work is part of the research program of the Foundation for Fundamental Research of Matter (FOM), which is part of The Netherlands Organisation for Scientific Research (NWO). We are grateful to M. van Huis  and J. Zaanen for critical reading of the manuscript, and M. Katsnelson for useful discussions.

\bibliographystyle{apsrev4-1}
\bibliography{Finite_size}

\begin{thebibliography}{26}%
\makeatletter
\providecommand \@ifxundefined [1]{%
 \@ifx{#1\undefined}
}%
\providecommand \@ifnum [1]{%
 \ifnum #1\expandafter \@firstoftwo
 \else \expandafter \@secondoftwo
 \fi
}%
\providecommand \@ifx [1]{%
 \ifx #1\expandafter \@firstoftwo
 \else \expandafter \@secondoftwo
 \fi
}%
\providecommand \natexlab [1]{#1}%
\providecommand \enquote  [1]{``#1''}%
\providecommand \bibnamefont  [1]{#1}%
\providecommand \bibfnamefont [1]{#1}%
\providecommand \citenamefont [1]{#1}%
\providecommand \href@noop [0]{\@secondoftwo}%
\providecommand \href [0]{\begingroup \@sanitize@url \@href}%
\providecommand \@href[1]{\@@startlink{#1}\@@href}%
\providecommand \@@href[1]{\endgroup#1\@@endlink}%
\providecommand \@sanitize@url [0]{\catcode `\\12\catcode `\$12\catcode
  `\&12\catcode `\#12\catcode `\^12\catcode `\_12\catcode `\%12\relax}%
\providecommand \@@startlink[1]{}%
\providecommand \@@endlink[0]{}%
\providecommand \url  [0]{\begingroup\@sanitize@url \@url }%
\providecommand \@url [1]{\endgroup\@href {#1}{\urlprefix }}%
\providecommand \urlprefix  [0]{URL }%
\providecommand \Eprint [0]{\href }%
\providecommand \doibase [0]{http://dx.doi.org/}%
\providecommand \selectlanguage [0]{\@gobble}%
\providecommand \bibinfo  [0]{\@secondoftwo}%
\providecommand \bibfield  [0]{\@secondoftwo}%
\providecommand \translation [1]{[#1]}%
\providecommand \BibitemOpen [0]{}%
\providecommand \bibitemStop [0]{}%
\providecommand \bibitemNoStop [0]{.\EOS\space}%
\providecommand \EOS [0]{\spacefactor3000\relax}%
\providecommand \BibitemShut  [1]{\csname bibitem#1\endcsname}%
\let\auto@bib@innerbib\@empty
\bibitem [{\citenamefont {Chen}\ \emph {et~al.}(2009)\citenamefont {Chen},
  \citenamefont {Cullen}, \citenamefont {Jang}, \citenamefont {Fuhrer},\ and\
  \citenamefont {Williams}}]{Chen2009}%
  \BibitemOpen
  \bibfield  {author} {\bibinfo {author} {\bibfnamefont {J.-H.}\ \bibnamefont
  {Chen}}, \bibinfo {author} {\bibfnamefont {W.~G.}\ \bibnamefont {Cullen}},
  \bibinfo {author} {\bibfnamefont {C.}~\bibnamefont {Jang}}, \bibinfo {author}
  {\bibfnamefont {M.~S.}\ \bibnamefont {Fuhrer}}, \ and\ \bibinfo {author}
  {\bibfnamefont {E.~D.}\ \bibnamefont {Williams}},\ }\href {\doibase
  10.1103/physrevlett.102.236805} {\bibfield  {journal} {\bibinfo  {journal}
  {Phys. Rev. Lett.}\ }\textbf {\bibinfo {volume} {102}},\ \bibinfo {pages}
  {236805} (\bibinfo {year} {2009})}\BibitemShut {NoStop}%
\bibitem [{\citenamefont {Zandiatashbar}\ \emph {et~al.}(2014)\citenamefont
  {Zandiatashbar}, \citenamefont {Lee}, \citenamefont {An}, \citenamefont
  {Lee}, \citenamefont {Mathew}, \citenamefont {Terrones}, \citenamefont
  {Hayashi}, \citenamefont {Picu}, \citenamefont {Hone},\ and\ \citenamefont
  {Koratkar}}]{Zandiatashbar2014}%
  \BibitemOpen
  \bibfield  {author} {\bibinfo {author} {\bibfnamefont {A.}~\bibnamefont
  {Zandiatashbar}}, \bibinfo {author} {\bibfnamefont {G.-H.}\ \bibnamefont
  {Lee}}, \bibinfo {author} {\bibfnamefont {S.~J.}\ \bibnamefont {An}},
  \bibinfo {author} {\bibfnamefont {S.}~\bibnamefont {Lee}}, \bibinfo {author}
  {\bibfnamefont {N.}~\bibnamefont {Mathew}}, \bibinfo {author} {\bibfnamefont
  {M.}~\bibnamefont {Terrones}}, \bibinfo {author} {\bibfnamefont
  {T.}~\bibnamefont {Hayashi}}, \bibinfo {author} {\bibfnamefont {C.~R.}\
  \bibnamefont {Picu}}, \bibinfo {author} {\bibfnamefont {J.}~\bibnamefont
  {Hone}}, \ and\ \bibinfo {author} {\bibfnamefont {N.}~\bibnamefont
  {Koratkar}},\ }\href@noop {} {\bibfield  {journal} {\bibinfo  {journal} {Nat.
  Commun.}\ }\textbf {\bibinfo {volume} {5}},\ \bibinfo {pages} {3186}
  (\bibinfo {year} {2014})}\BibitemShut {NoStop}%
\bibitem [{\citenamefont {Lopez-Polin}\ \emph {et~al.}(2014)\citenamefont
  {Lopez-Polin}, \citenamefont {Gomez-Navarro}, \citenamefont {Parente},
  \citenamefont {Guinea}, \citenamefont {Katsnelson}, \citenamefont
  {Pérez-Murano},\ and\ \citenamefont {Gómez-Herrero}}]{Lapez-2014}%
  \BibitemOpen
  \bibfield  {author} {\bibinfo {author} {\bibfnamefont {G.}~\bibnamefont
  {Lopez-Polin}}, \bibinfo {author} {\bibfnamefont {C.}~\bibnamefont
  {Gomez-Navarro}}, \bibinfo {author} {\bibfnamefont {V.}~\bibnamefont
  {Parente}}, \bibinfo {author} {\bibfnamefont {F.}~\bibnamefont {Guinea}},
  \bibinfo {author} {\bibfnamefont {M.}~\bibnamefont {Katsnelson}}, \bibinfo
  {author} {\bibfnamefont {F.}~\bibnamefont {Pérez-Murano}}, \ and\ \bibinfo
  {author} {\bibfnamefont {J.}~\bibnamefont {Gómez-Herrero}},\ }\href
  {\doibase 10.1038/nphys3183} {\bibfield  {journal} {\bibinfo  {journal} {Nat.
  Phys.}\ }\textbf {\bibinfo {volume} {11}},\ \bibinfo {pages} {26} (\bibinfo
  {year} {2014})}\BibitemShut {NoStop}%
\bibitem [{\citenamefont {Grantab}\ \emph {et~al.}(2010)\citenamefont
  {Grantab}, \citenamefont {Shenoy},\ and\ \citenamefont
  {Ruoff}}]{Grantab2010}%
  \BibitemOpen
  \bibfield  {author} {\bibinfo {author} {\bibfnamefont {R.}~\bibnamefont
  {Grantab}}, \bibinfo {author} {\bibfnamefont {V.~B.}\ \bibnamefont {Shenoy}},
  \ and\ \bibinfo {author} {\bibfnamefont {R.~S.}\ \bibnamefont {Ruoff}},\
  }\href {\doibase 10.1126/science.1196893} {\bibfield  {journal} {\bibinfo
  {journal} {Science}\ }\textbf {\bibinfo {volume} {330}},\ \bibinfo {pages}
  {946} (\bibinfo {year} {2010})}\BibitemShut {NoStop}%
\bibitem [{\citenamefont {Skowron}\ \emph {et~al.}(2015)\citenamefont
  {Skowron}, \citenamefont {Lebedeva}, \citenamefont {Popov},\ and\
  \citenamefont {Bichoutskaia}}]{Skowron2015}%
  \BibitemOpen
  \bibfield  {author} {\bibinfo {author} {\bibfnamefont {S.~T.}\ \bibnamefont
  {Skowron}}, \bibinfo {author} {\bibfnamefont {I.~V.}\ \bibnamefont
  {Lebedeva}}, \bibinfo {author} {\bibfnamefont {A.~M.}\ \bibnamefont {Popov}},
  \ and\ \bibinfo {author} {\bibfnamefont {E.}~\bibnamefont {Bichoutskaia}},\
  }\href {\doibase 10.1039/c4cs00499j} {\bibfield  {journal} {\bibinfo
  {journal} {Chem. Soc. Rev.}\ }\textbf {\bibinfo {volume} {44}},\ \bibinfo
  {pages} {3143} (\bibinfo {year} {2015})}\BibitemShut {NoStop}%
\bibitem [{\citenamefont {Fasolino}\ \emph {et~al.}(2007)\citenamefont
  {Fasolino}, \citenamefont {Los},\ and\ \citenamefont
  {Katsnelson}}]{Fasolino2007}%
  \BibitemOpen
  \bibfield  {author} {\bibinfo {author} {\bibfnamefont {A.}~\bibnamefont
  {Fasolino}}, \bibinfo {author} {\bibfnamefont {J.~H.}\ \bibnamefont {Los}}, \
  and\ \bibinfo {author} {\bibfnamefont {M.~I.}\ \bibnamefont {Katsnelson}},\
  }\href {\doibase 10.1038/nmat2011} {\bibfield  {journal} {\bibinfo  {journal}
  {Nat. Mater.}\ }\textbf {\bibinfo {volume} {6}},\ \bibinfo {pages} {858}
  (\bibinfo {year} {2007})}\BibitemShut {NoStop}%
\bibitem [{\citenamefont {Banhart}\ \emph {et~al.}(2011)\citenamefont
  {Banhart}, \citenamefont {Kotakoski},\ and\ \citenamefont
  {Krasheninnikov}}]{banhart-review-2011}%
  \BibitemOpen
  \bibfield  {author} {\bibinfo {author} {\bibfnamefont {F.}~\bibnamefont
  {Banhart}}, \bibinfo {author} {\bibfnamefont {J.}~\bibnamefont {Kotakoski}},
  \ and\ \bibinfo {author} {\bibfnamefont {A.~V.}\ \bibnamefont
  {Krasheninnikov}},\ }\href {\doibase 10.1021/nn102598m} {\bibfield  {journal}
  {\bibinfo  {journal} {ACS Nano}\ }\textbf {\bibinfo {volume} {5}},\ \bibinfo
  {pages} {26} (\bibinfo {year} {2011})}\BibitemShut {NoStop}%
\bibitem [{\citenamefont {Zhang}\ \emph {et~al.}(2014)\citenamefont {Zhang},
  \citenamefont {Li},\ and\ \citenamefont {Gao}}]{Zhang2014}%
  \BibitemOpen
  \bibfield  {author} {\bibinfo {author} {\bibfnamefont {T.}~\bibnamefont
  {Zhang}}, \bibinfo {author} {\bibfnamefont {X.}~\bibnamefont {Li}}, \ and\
  \bibinfo {author} {\bibfnamefont {H.}~\bibnamefont {Gao}},\ }\href {\doibase
  10.1016/j.jmps.2014.02.005} {\bibfield  {journal} {\bibinfo  {journal} {J.
  Mech. and Phys. Solids}\ }\textbf {\bibinfo {volume} {67}},\ \bibinfo {pages}
  {2} (\bibinfo {year} {2014})}\BibitemShut {NoStop}%
\bibitem [{SOM()}]{SOM}%
  \BibitemOpen
  \href@noop {} {}\bibinfo {note} {See Supplementary Online Material, which
  includes (1) detailed derivations of (1+1)D and (2+1)D analytical models; (2)
  computational details for numerical simulation of graphene; and (3)
  dislocation separation potential}\BibitemShut {NoStop}%
\bibitem [{\citenamefont {Jain}\ \emph {et~al.}(2015)\citenamefont {Jain},
  \citenamefont {Barkema}, \citenamefont {Mousseau}, \citenamefont {Fang},\
  and\ \citenamefont {van Huis}}]{Sandeep2015}%
  \BibitemOpen
  \bibfield  {author} {\bibinfo {author} {\bibfnamefont {S.~K.}\ \bibnamefont
  {Jain}}, \bibinfo {author} {\bibfnamefont {G.~T.}\ \bibnamefont {Barkema}},
  \bibinfo {author} {\bibfnamefont {N.}~\bibnamefont {Mousseau}}, \bibinfo
  {author} {\bibfnamefont {C.-M.}\ \bibnamefont {Fang}}, \ and\ \bibinfo
  {author} {\bibfnamefont {M.~A.}\ \bibnamefont {van Huis}},\ }\href {\doibase
  10.1021/acs.jpcc.5b01905} {\bibfield  {journal} {\bibinfo  {journal} {J.
  Phys. Chem. C}\ }\textbf {\bibinfo {volume} {119}},\ \bibinfo {pages} {9646}
  (\bibinfo {year} {2015})}\BibitemShut {NoStop}%
\bibitem [{\citenamefont {Los}\ \emph {et~al.}(2015)\citenamefont {Los},
  \citenamefont {Zakharchenko}, \citenamefont {Katsnelson},\ and\ \citenamefont
  {Fasolino}}]{Los2015}%
  \BibitemOpen
  \bibfield  {author} {\bibinfo {author} {\bibfnamefont {J.~H.}\ \bibnamefont
  {Los}}, \bibinfo {author} {\bibfnamefont {K.~V.}\ \bibnamefont
  {Zakharchenko}}, \bibinfo {author} {\bibfnamefont {M.~I.}\ \bibnamefont
  {Katsnelson}}, \ and\ \bibinfo {author} {\bibfnamefont {A.}~\bibnamefont
  {Fasolino}},\ }\href {\doibase 10.1103/physrevb.91.045415} {\bibfield
  {journal} {\bibinfo  {journal} {Phys. Rev. B}\ }\textbf {\bibinfo {volume}
  {91}},\ \bibinfo {pages} {045415} (\bibinfo {year} {2015})}\BibitemShut
  {NoStop}%
\bibitem [{\citenamefont {Zakharchenko}\ \emph {et~al.}(2011)\citenamefont
  {Zakharchenko}, \citenamefont {Fasolino}, \citenamefont {Los},\ and\
  \citenamefont {Katsnelson}}]{Zakharchenko2011}%
  \BibitemOpen
  \bibfield  {author} {\bibinfo {author} {\bibfnamefont {K.~V.}\ \bibnamefont
  {Zakharchenko}}, \bibinfo {author} {\bibfnamefont {A.}~\bibnamefont
  {Fasolino}}, \bibinfo {author} {\bibfnamefont {J.~H.}\ \bibnamefont {Los}}, \
  and\ \bibinfo {author} {\bibfnamefont {M.~I.}\ \bibnamefont {Katsnelson}},\
  }\href {\doibase 10.1088/0953-8984/23/20/202202} {\bibfield  {journal}
  {\bibinfo  {journal} {J. Phys. Condens. Matter}\ }\textbf {\bibinfo {volume}
  {23}},\ \bibinfo {pages} {202202} (\bibinfo {year} {2011})}\BibitemShut
  {NoStop}%
\bibitem [{\citenamefont {Seung}\ and\ \citenamefont
  {Nelson}(1988)}]{Seung1988}%
  \BibitemOpen
  \bibfield  {author} {\bibinfo {author} {\bibfnamefont {H.~S.}\ \bibnamefont
  {Seung}}\ and\ \bibinfo {author} {\bibfnamefont {D.~R.}\ \bibnamefont
  {Nelson}},\ }\href {\doibase 10.1103/physreva.38.1005} {\bibfield  {journal}
  {\bibinfo  {journal} {Physical Review A}\ }\textbf {\bibinfo {volume} {38}},\
  \bibinfo {pages} {1005} (\bibinfo {year} {1988})}\BibitemShut {NoStop}%
\bibitem [{\citenamefont {Carlsson}\ \emph {et~al.}(2011)\citenamefont
  {Carlsson}, \citenamefont {Ghiringhelli},\ and\ \citenamefont
  {Fasolino}}]{Carlsson2011}%
  \BibitemOpen
  \bibfield  {author} {\bibinfo {author} {\bibfnamefont {J.~M.}\ \bibnamefont
  {Carlsson}}, \bibinfo {author} {\bibfnamefont {L.~M.}\ \bibnamefont
  {Ghiringhelli}}, \ and\ \bibinfo {author} {\bibfnamefont {A.}~\bibnamefont
  {Fasolino}},\ }\href {\doibase 10.1103/physrevb.84.165423} {\bibfield
  {journal} {\bibinfo  {journal} {Physical Review B}\ }\textbf {\bibinfo
  {volume} {84}},\ \bibinfo {pages} {165423} (\bibinfo {year}
  {2011})}\BibitemShut {NoStop}%
\bibitem [{\citenamefont {Ertekin}\ \emph {et~al.}(2009)\citenamefont
  {Ertekin}, \citenamefont {Chrzan},\ and\ \citenamefont {Daw}}]{Ertekin2009}%
  \BibitemOpen
  \bibfield  {author} {\bibinfo {author} {\bibfnamefont {E.}~\bibnamefont
  {Ertekin}}, \bibinfo {author} {\bibfnamefont {D.~C.}\ \bibnamefont {Chrzan}},
  \ and\ \bibinfo {author} {\bibfnamefont {M.~S.}\ \bibnamefont {Daw}},\ }\href
  {\doibase 10.1103/physrevb.79.155421} {\bibfield  {journal} {\bibinfo
  {journal} {Phys. Rev. B}\ }\textbf {\bibinfo {volume} {79}},\ \bibinfo
  {pages} {155421} (\bibinfo {year} {2009})}\BibitemShut {NoStop}%
\bibitem [{\citenamefont {Hashimoto}\ \emph {et~al.}(2004)\citenamefont
  {Hashimoto}, \citenamefont {Suenaga}, \citenamefont {Gloter}, \citenamefont
  {Urita},\ and\ \citenamefont {Iijima}}]{Hashimoto2004}%
  \BibitemOpen
  \bibfield  {author} {\bibinfo {author} {\bibfnamefont {A.}~\bibnamefont
  {Hashimoto}}, \bibinfo {author} {\bibfnamefont {K.}~\bibnamefont {Suenaga}},
  \bibinfo {author} {\bibfnamefont {A.}~\bibnamefont {Gloter}}, \bibinfo
  {author} {\bibfnamefont {K.}~\bibnamefont {Urita}}, \ and\ \bibinfo {author}
  {\bibfnamefont {S.}~\bibnamefont {Iijima}},\ }\href {\doibase
  10.1038/nature02817} {\bibfield  {journal} {\bibinfo  {journal} {Nature}\
  }\textbf {\bibinfo {volume} {430}},\ \bibinfo {pages} {870} (\bibinfo {year}
  {2004})}\BibitemShut {NoStop}%
\bibitem [{\citenamefont {Meyer}\ \emph {et~al.}(2008)\citenamefont {Meyer},
  \citenamefont {Kisielowski}, \citenamefont {Erni}, \citenamefont {Rossell},
  \citenamefont {Crommie},\ and\ \citenamefont {Zettl}}]{Meyer2008}%
  \BibitemOpen
  \bibfield  {author} {\bibinfo {author} {\bibfnamefont {J.~C.}\ \bibnamefont
  {Meyer}}, \bibinfo {author} {\bibfnamefont {C.}~\bibnamefont {Kisielowski}},
  \bibinfo {author} {\bibfnamefont {R.}~\bibnamefont {Erni}}, \bibinfo {author}
  {\bibfnamefont {M.~D.}\ \bibnamefont {Rossell}}, \bibinfo {author}
  {\bibfnamefont {M.~F.}\ \bibnamefont {Crommie}}, \ and\ \bibinfo {author}
  {\bibfnamefont {A.}~\bibnamefont {Zettl}},\ }\href {\doibase
  10.1021/nl801386m} {\bibfield  {journal} {\bibinfo  {journal} {Nano Lett.}\
  }\textbf {\bibinfo {volume} {8}},\ \bibinfo {pages} {3582} (\bibinfo {year}
  {2008})}\BibitemShut {NoStop}%
\bibitem [{\citenamefont {Kotakoski}\ \emph {et~al.}(2011)\citenamefont
  {Kotakoski}, \citenamefont {Krasheninnikov}, \citenamefont {Kaiser},\ and\
  \citenamefont {Meyer}}]{Kotakoski2011}%
  \BibitemOpen
  \bibfield  {author} {\bibinfo {author} {\bibfnamefont {J.}~\bibnamefont
  {Kotakoski}}, \bibinfo {author} {\bibfnamefont {A.~V.}\ \bibnamefont
  {Krasheninnikov}}, \bibinfo {author} {\bibfnamefont {U.}~\bibnamefont
  {Kaiser}}, \ and\ \bibinfo {author} {\bibfnamefont {J.~C.}\ \bibnamefont
  {Meyer}},\ }\href {\doibase 10.1103/physrevlett.106.105505} {\bibfield
  {journal} {\bibinfo  {journal} {Phys. Rev. Lett.}\ }\textbf {\bibinfo
  {volume} {106}},\ \bibinfo {pages} {105505} (\bibinfo {year}
  {2011})}\BibitemShut {NoStop}%
\bibitem [{\citenamefont {Huang}\ \emph {et~al.}(2011)\citenamefont {Huang},
  \citenamefont {Ruiz-Vargas}, \citenamefont {van~der Zande}, \citenamefont
  {Whitney}, \citenamefont {Levendorf}, \citenamefont {Kevek}, \citenamefont
  {Garg}, \citenamefont {Alden}, \citenamefont {Hustedt}, \citenamefont {Zhu},\
  and\ \citenamefont {et~al.}}]{Huang2011}%
  \BibitemOpen
  \bibfield  {author} {\bibinfo {author} {\bibfnamefont {P.~Y.}\ \bibnamefont
  {Huang}}, \bibinfo {author} {\bibfnamefont {C.~S.}\ \bibnamefont
  {Ruiz-Vargas}}, \bibinfo {author} {\bibfnamefont {A.~M.}\ \bibnamefont
  {van~der Zande}}, \bibinfo {author} {\bibfnamefont {W.~S.}\ \bibnamefont
  {Whitney}}, \bibinfo {author} {\bibfnamefont {M.~P.}\ \bibnamefont
  {Levendorf}}, \bibinfo {author} {\bibfnamefont {J.~W.}\ \bibnamefont
  {Kevek}}, \bibinfo {author} {\bibfnamefont {S.}~\bibnamefont {Garg}},
  \bibinfo {author} {\bibfnamefont {J.~S.}\ \bibnamefont {Alden}}, \bibinfo
  {author} {\bibfnamefont {C.~J.}\ \bibnamefont {Hustedt}}, \bibinfo {author}
  {\bibfnamefont {Y.}~\bibnamefont {Zhu}}, \ and\ \bibinfo {author}
  {\bibnamefont {et~al.}},\ }\href {\doibase 10.1038/nature09718} {\bibfield
  {journal} {\bibinfo  {journal} {Nature}\ }\textbf {\bibinfo {volume} {469}},\
  \bibinfo {pages} {389} (\bibinfo {year} {2011})}\BibitemShut {NoStop}%
\bibitem [{\citenamefont {Yazyev}\ and\ \citenamefont
  {Chen}(2014)}]{Yazyev2014}%
  \BibitemOpen
  \bibfield  {author} {\bibinfo {author} {\bibfnamefont {O.~V.}\ \bibnamefont
  {Yazyev}}\ and\ \bibinfo {author} {\bibfnamefont {Y.~P.}\ \bibnamefont
  {Chen}},\ }\href {\doibase 10.1038/nnano.2014.166} {\bibfield  {journal}
  {\bibinfo  {journal} {Nat. Nanotechnol.}\ }\textbf {\bibinfo {volume} {9}},\
  \bibinfo {pages} {755} (\bibinfo {year} {2014})}\BibitemShut {NoStop}%
\bibitem [{\citenamefont {Tsen}\ \emph {et~al.}(2013)\citenamefont {Tsen},
  \citenamefont {Brown}, \citenamefont {Havener},\ and\ \citenamefont
  {Park}}]{Tsen2013}%
  \BibitemOpen
  \bibfield  {author} {\bibinfo {author} {\bibfnamefont {A.~W.}\ \bibnamefont
  {Tsen}}, \bibinfo {author} {\bibfnamefont {L.}~\bibnamefont {Brown}},
  \bibinfo {author} {\bibfnamefont {R.~W.}\ \bibnamefont {Havener}}, \ and\
  \bibinfo {author} {\bibfnamefont {J.}~\bibnamefont {Park}},\ }\href {\doibase
  10.1021/ar300190z} {\bibfield  {journal} {\bibinfo  {journal} {Acc. Chem.
  Res.}\ }\textbf {\bibinfo {volume} {46}},\ \bibinfo {pages} {2286} (\bibinfo
  {year} {2013})}\BibitemShut {NoStop}%
\bibitem [{\citenamefont {Xu}\ \emph {et~al.}(2015)\citenamefont {Xu},
  \citenamefont {Wang}, \citenamefont {Liu}, \citenamefont {Chen},
  \citenamefont {Guo}, \citenamefont {Kang}, \citenamefont {Ma}, \citenamefont
  {Cheng},\ and\ \citenamefont {Ren}}]{Xu2015}%
  \BibitemOpen
  \bibfield  {author} {\bibinfo {author} {\bibfnamefont {C.}~\bibnamefont
  {Xu}}, \bibinfo {author} {\bibfnamefont {L.}~\bibnamefont {Wang}}, \bibinfo
  {author} {\bibfnamefont {Z.}~\bibnamefont {Liu}}, \bibinfo {author}
  {\bibfnamefont {L.}~\bibnamefont {Chen}}, \bibinfo {author} {\bibfnamefont
  {J.}~\bibnamefont {Guo}}, \bibinfo {author} {\bibfnamefont {N.}~\bibnamefont
  {Kang}}, \bibinfo {author} {\bibfnamefont {X.-L.}\ \bibnamefont {Ma}},
  \bibinfo {author} {\bibfnamefont {H.-M.}\ \bibnamefont {Cheng}}, \ and\
  \bibinfo {author} {\bibfnamefont {W.}~\bibnamefont {Ren}},\ }\href {\doibase
  10.1038/nmat4374} {\bibfield  {journal} {\bibinfo  {journal} {Nat. Mater.}\
  }\textbf {\bibinfo {volume} {14}},\ \bibinfo {pages} {1135} (\bibinfo {year}
  {2015})}\BibitemShut {NoStop}%
\bibitem [{\citenamefont {Vogt}\ \emph {et~al.}(2012)\citenamefont {Vogt},
  \citenamefont {De~Padova}, \citenamefont {Quaresima}, \citenamefont {Avila},
  \citenamefont {Frantzeskakis}, \citenamefont {Asensio}, \citenamefont
  {Resta}, \citenamefont {Ealet},\ and\ \citenamefont {Le~Lay}}]{Vogt2012}%
  \BibitemOpen
  \bibfield  {author} {\bibinfo {author} {\bibfnamefont {P.}~\bibnamefont
  {Vogt}}, \bibinfo {author} {\bibfnamefont {P.}~\bibnamefont {De~Padova}},
  \bibinfo {author} {\bibfnamefont {C.}~\bibnamefont {Quaresima}}, \bibinfo
  {author} {\bibfnamefont {J.}~\bibnamefont {Avila}}, \bibinfo {author}
  {\bibfnamefont {E.}~\bibnamefont {Frantzeskakis}}, \bibinfo {author}
  {\bibfnamefont {M.~C.}\ \bibnamefont {Asensio}}, \bibinfo {author}
  {\bibfnamefont {A.}~\bibnamefont {Resta}}, \bibinfo {author} {\bibfnamefont
  {B.}~\bibnamefont {Ealet}}, \ and\ \bibinfo {author} {\bibfnamefont
  {G.}~\bibnamefont {Le~Lay}},\ }\href {\doibase
  10.1103/physrevlett.108.155501} {\bibfield  {journal} {\bibinfo  {journal}
  {Phys. Rev. Lett.}\ }\textbf {\bibinfo {volume} {108}},\ \bibinfo {pages}
  {155501} (\bibinfo {year} {2012})}\BibitemShut {NoStop}%
\bibitem [{\citenamefont {Feng}\ \emph {et~al.}(2012)\citenamefont {Feng},
  \citenamefont {Ding}, \citenamefont {Meng}, \citenamefont {Yao},
  \citenamefont {He}, \citenamefont {Cheng}, \citenamefont {Chen},\ and\
  \citenamefont {Wu}}]{Feng2012}%
  \BibitemOpen
  \bibfield  {author} {\bibinfo {author} {\bibfnamefont {B.}~\bibnamefont
  {Feng}}, \bibinfo {author} {\bibfnamefont {Z.}~\bibnamefont {Ding}}, \bibinfo
  {author} {\bibfnamefont {S.}~\bibnamefont {Meng}}, \bibinfo {author}
  {\bibfnamefont {Y.}~\bibnamefont {Yao}}, \bibinfo {author} {\bibfnamefont
  {X.}~\bibnamefont {He}}, \bibinfo {author} {\bibfnamefont {P.}~\bibnamefont
  {Cheng}}, \bibinfo {author} {\bibfnamefont {L.}~\bibnamefont {Chen}}, \ and\
  \bibinfo {author} {\bibfnamefont {K.}~\bibnamefont {Wu}},\ }\href {\doibase
  10.1021/nl301047g} {\bibfield  {journal} {\bibinfo  {journal} {Nano Lett.}\
  }\textbf {\bibinfo {volume} {12}},\ \bibinfo {pages} {3507} (\bibinfo {year}
  {2012})}\BibitemShut {NoStop}%
\bibitem [{\citenamefont {Bianco}\ \emph {et~al.}(2013)\citenamefont {Bianco},
  \citenamefont {Butler}, \citenamefont {Jiang}, \citenamefont {Restrepo},
  \citenamefont {Windl},\ and\ \citenamefont {Goldberger}}]{Bianco2013}%
  \BibitemOpen
  \bibfield  {author} {\bibinfo {author} {\bibfnamefont {E.}~\bibnamefont
  {Bianco}}, \bibinfo {author} {\bibfnamefont {S.}~\bibnamefont {Butler}},
  \bibinfo {author} {\bibfnamefont {S.}~\bibnamefont {Jiang}}, \bibinfo
  {author} {\bibfnamefont {O.~D.}\ \bibnamefont {Restrepo}}, \bibinfo {author}
  {\bibfnamefont {W.}~\bibnamefont {Windl}}, \ and\ \bibinfo {author}
  {\bibfnamefont {J.~E.}\ \bibnamefont {Goldberger}},\ }\href {\doibase
  10.1021/nn4009406} {\bibfield  {journal} {\bibinfo  {journal} {ACS Nano}\
  }\textbf {\bibinfo {volume} {7}},\ \bibinfo {pages} {4414} (\bibinfo {year}
  {2013})}\BibitemShut {NoStop}%
\bibitem [{\citenamefont {Zhu}\ \emph {et~al.}(2015)\citenamefont {Zhu},
  \citenamefont {Chen}, \citenamefont {Xu}, \citenamefont {Gao}, \citenamefont
  {Guan}, \citenamefont {Liu}, \citenamefont {Qian}, \citenamefont {Zhang},\
  and\ \citenamefont {Jia}}]{Zhu2015}%
  \BibitemOpen
  \bibfield  {author} {\bibinfo {author} {\bibfnamefont {F.-f.}\ \bibnamefont
  {Zhu}}, \bibinfo {author} {\bibfnamefont {W.-j.}\ \bibnamefont {Chen}},
  \bibinfo {author} {\bibfnamefont {Y.}~\bibnamefont {Xu}}, \bibinfo {author}
  {\bibfnamefont {C.-l.}\ \bibnamefont {Gao}}, \bibinfo {author} {\bibfnamefont
  {D.-d.}\ \bibnamefont {Guan}}, \bibinfo {author} {\bibfnamefont {C.-h.}\
  \bibnamefont {Liu}}, \bibinfo {author} {\bibfnamefont {D.}~\bibnamefont
  {Qian}}, \bibinfo {author} {\bibfnamefont {S.-C.}\ \bibnamefont {Zhang}}, \
  and\ \bibinfo {author} {\bibfnamefont {J.-f.}\ \bibnamefont {Jia}},\ }\href
  {\doibase 10.1038/nmat4384} {\bibfield  {journal} {\bibinfo  {journal} {Nat.
  Mater.}\ }\textbf {\bibinfo {volume} {14}},\ \bibinfo {pages} {1020}
  (\bibinfo {year} {2015})}\BibitemShut {NoStop}%
\end{thebibliography}%

\vspace{5cm}

\pagebreak

\onecolumngrid

\begin{center}
{\bf Supplementary Materials for ``\emph{Boundaries determine the formation energies of lattice defects in two-dimensional buckled materials}"} \\

Sandeep K. Jain$^1$, Vladimir Juri\v ci\' c$^{2,1}$ and Gerard T. Barkema$^{1,3}$\\

$^{1}$\emph{Institute for Theoretical Physics, Universiteit Utrecht, Leuvenlaan 4, 3584 CE Utrecht, The Netherlands} \\

$^{2}$\emph{Nordita,  Center for Quantum Materials,  KTH Royal Institute of Technology and Stockholm University, Roslagstullsbacken 23,  10691 Stockholm, Sweden}

$^{3}$\emph{Instituut-Lorentz for Theoretical Physics, Universiteit Leiden, P.O. Box 9506, 2300 RA Leiden, The Netherlands}
\end{center}

$\bullet$ \emph{\bf Effective (1+1)D and (2+1)D models.}

As discussed in the main text, effective models are developed to
understand the effect of boundary conditions on the formation energy of
defects in (1+1) and (2+1) dimensions.\\

1. {\bf (1+1)D model.} \\
In the (1+1)D model, a linear
string of atoms is bonded by harmonic springs with unit ideal length,
and neighboring bonds prefer to be aligned. The system has periodic
boundary conditions, with a periodic length $L$.  The ground state is
thus a straight linear, periodic set of atoms in which atom $i=1\dots N$
has the coordinates $\vec{r}_i=(i,0)$; and the periodic length is $L=N$ as shown in Fig.\ 1(a) in the main text.
The Hamiltonian is
\begin{align}
H=E_{core}+\lambda\sum_{i}(r_{i}-1)^2+\kappa\sum_{i}(\phi_{i+1}-\phi_{i})^2-f\left(\sum_{i}r_{i}\cos(\phi_{i})-L\right).
\end{align}
The parameters of the Hamiltonian are defined below Eq. (1) in the main text.

In this system, a defect is introduced as a single atom with its bonds that prefer
to make an angle $\Delta\phi \neq 0$. Below, we present an analytic solution
of the energy-minimized positions of the atoms, and in case of FF boundary
conditions the periodic length.

\underline{FF boundary solution (at relaxed boundaries)}:
With an FF boundary, the net force on the boundary is zero and the energy is minimized for
\begin{align}
\frac{\partial E}{\partial r_{i}}=2\lambda(r_{i}-1)=0,
\end{align}
\begin{align}
\frac{\partial E}{\partial \phi_{i}}=-2\kappa(\phi_{i+1}-\phi_{i})+2\kappa(\phi_{i}-\phi_{i-1})=0.
\end{align}

Going to the continuum limit (lattice spacing tends to zero), with $i \rightarrow \rho$,
the last equation becomes
\begin{equation}
2\kappa\frac{\partial^{2} \phi(\rho)}{\partial \rho^{2}}=0,
\end{equation}
and  the solutions read
\begin{align}
r(\rho)=1,
\end{align}
\begin{align}
\phi(\rho)=\phi_{0}(-1+\frac{\rho}{N}).
\end{align}

Using these solutions, we obtain a finite-size correction of the energy $E\sim N^{-1}$.
We verified this analytical result with numerical simulation data as shown in Fig.\ 1(b) in the main text. The periodic length in this case is
\begin{align}
L=N\left(1-\frac{\phi_{0}^{2}}{6}\right).
\end{align}

\underline{DF solution (at fixed boundaries)}:
With a DF boundary, a net force is acting on the ends of the string to keep the density of atoms fixed. Therefore, the energy is minimized for
\begin{align}\label{DF-1-r-cont}
\frac{\partial E}{\partial r_{i}}=2\lambda(r_{i}-1)-f\cos(\phi_{i})=0,
\end{align}
\begin{align}
\frac{\partial E}{\partial \phi_{i}}=-2\kappa(\phi_{i+1}-\phi_{i})+2\kappa(\phi_{i}-\phi_{i-1})+fr_{i}\sin(\phi_{i})=0.
\end{align}
For small values of $\phi_{i}$, we use $\sin(\phi_{i})\approx\phi_{i}$, to obtain
\begin{align}
\frac{\partial E}{\partial \phi_{i}}=2\kappa(2\phi_{i}-\phi_{i+1}-\phi_{i-1})+fr_{i}\phi_{i}=0.
\end{align}
This equation in the continuum limit($i \rightarrow\rho$) then reads
\begin{align}\label{DF-1-phi-cont}
2\kappa\frac{\partial^{2} \phi(\rho)}{\partial \rho^{2}}-fr({\rho})\phi({\rho})=0.
\end{align}
For small values of $\phi_{i}$,  $\cos(\phi_{i})\approx 1$ and in continuum limit $i \rightarrow \rho$, the solutions of  Eqs.\ (\ref{DF-1-r-cont}) and (\ref{DF-1-phi-cont}) read
\begin{align}
r(\rho)=1+\frac{f}{2\lambda}
\end{align}
\begin{align}
\phi(\rho)=\phi_{0} e^{-\alpha \rho},
\end{align}
with
\begin{equation}
\alpha=\sqrt{\frac{fr(\rho)}{2\kappa}}.
\end{equation}

The force required to make the atomic density equal to crystalline density is obtained from the condition of
the constant length of the string
\begin{align}
\sum\limits_{i}r_{i}\cos(\phi_{i})=x_{N}-x_{0}=N.
\end{align}
In the continuum limit, $i \rightarrow \rho$
\begin{align}
\int_{0}^{N} d\rho \, r(\rho)\,\cos\phi(\rho)=N.
\end{align}
After substituting the solutions for $r(\rho)$ and $\phi(\rho)$ and using that for $\phi<<1$, $\cos\phi\approx1-\frac{\phi^2}{2}$, we obtain
\begin{align}
(1+\frac{f}{2\lambda})\int_{0}^{N} d\rho \left(1-\frac{\phi_{0}^2}{2}e^{-2\alpha \rho}\right)=N,
\end{align}
which after performing the integral yields
\begin{align}
\frac{fN}{2\lambda}-(1+\frac{f}{2\lambda})\frac{\phi_{0}^{2}}{4\alpha}=0.
\end{align}
Substituting $\alpha=\sqrt{\frac{f\,r(\rho)}{2\kappa}}$ and $r(\rho)=1+\frac{f}{2\lambda}\approx 1$, we obtain
\begin{align}
f\sim N^{-2/3}.
\end{align}
Using this result, the finite-size energy for the given solutions scales as $E\sim N^{-1/3}$. We verified
this analytical result with numerical simulation data as shown in Fig.\ 1(b).

2. {\bf (2+1)D model} 

The (1+1)D model is then generalized to (2+1)D by imposing rotational symmetry,
and constraining the model such that the structure is flat at its perimeter. The corresponding Hamiltonian reads
\begin{align}\label{Ham-2+1}
H=E_{core}+\lambda\sum_{i}i(r_{i}-1)^2+\kappa\sum_{i}i(\phi_{i+1}-\phi_{i})^2-f\left(\sum_{i}r_{i}\cos(\phi_{i})-L\right).
\end{align}
\underline{FF boundary solution (no pulling on the perimeter)}:
At the FF boundary, the net force on the boundary is zero and the energy is minimized for
\begin{align}
\frac{\partial E}{\partial r_{i}}=2i\lambda(r_{i}-1)=0,
\end{align}
\begin{align}
\frac{\partial E}{\partial \phi_{i}}=-2i\kappa(\phi_{i+1}-\phi_{i})+2(i-1)\kappa(\phi_{i}-\phi_{i-1})=0,
\end{align}
which in the continuum limit reads
\begin{equation}
2\rho\kappa\frac{\partial^{2} \phi({\rho})}{\partial \rho^{2}}+2\kappa\frac{\partial \phi({\rho})}{\partial \rho}=0.
\end{equation}
The solutions of the above equations read
\begin{align}
r(\rho)=1,
\end{align}
\begin{align}
\phi(\rho)=\phi_{0}\left(1-\frac{\log(2\rho+1)}{\log(2N+1)}\right).
\end{align}
The energy corresponding to the above solutions of $r(\rho)$ and $\phi(\rho)$ scales with the system size as
$E\sim {1 \over \log(N)}$. This analytical result is in very good agreement with numerical simulation
data as shown in Fig.\ 1(c). The periodic length in this case is
\begin{align}
L=N-\frac{\phi_{0}^{2}}{2}\left[-\frac{1}{2}-\frac{1}{\log(2N+1)}+\frac{2N}{\log(2N+1)^2}\right].
\end{align}

\underline{DF solution (at fixed boundaries)}: At the DF boundary, a net force is acting on the ends of the string to keep the density of atoms fixed. Hamiltonian (\ref{Ham-2+1}) then yields
\begin{align}
\frac{\partial E}{\partial r_{i}}=2i\lambda(r_{i}-1)-f\cos(\phi_{i})=0
\end{align}
\begin{align}
\frac{\partial E}{\partial \phi_{i}}=-2i\kappa(\phi_{i+1}-\phi_{i})+2(i-1)\kappa(\phi_{i}-\phi_{i-1})+fr_{i}\sin(\phi_{i})=0.
\end{align}
In the continuum limit, for small $\phi\ll 1$, the last equation becomes
\begin{align}
2\rho\kappa\frac{\partial^{2} \phi({\rho})}{\partial \rho^{2}}+2\kappa\frac{\partial \phi_{\rho}}{\partial \rho}-fr(\rho)\phi({\rho})=0.
\end{align}
The corresponding solutions in the continuum limit read
\begin{align}
r(\rho)=1+\frac{f}{2\rho\lambda},
\end{align}
\begin{align}
\phi(\rho)=K_{0}\left(\sqrt{\frac{2f\rho}{\kappa}}\right),
\end{align}
where $K_0(x)$ is the modified Bessel function of the second kind.

To calculate the force we used the same equation as in (1+1)D model, since the (2+1)D model is rotationally symmetric. In this case,
using that
\begin{align}
\int_{0}^{\infty}d\rho\, \left[K_0\left(\sqrt{\frac{2f\rho}{\kappa}}\right)\right]^2=\frac{\kappa}{2f},
\end{align}
 we obtain that force scales as
$f\sim$ $ 1\over \sqrt{\log(N)}$ whereas energy scales as $E\sim$ $ {1 \over \log(N)}$ with a constant offset. This simple (2+1)D model gives the offset in energy for two different boundaries in the thermodynamic limit which is also shown in  graphene samples with defects via computer simulations.
\\

\begin{figure}
\begin{center}
\includegraphics[width=1.0\textwidth]{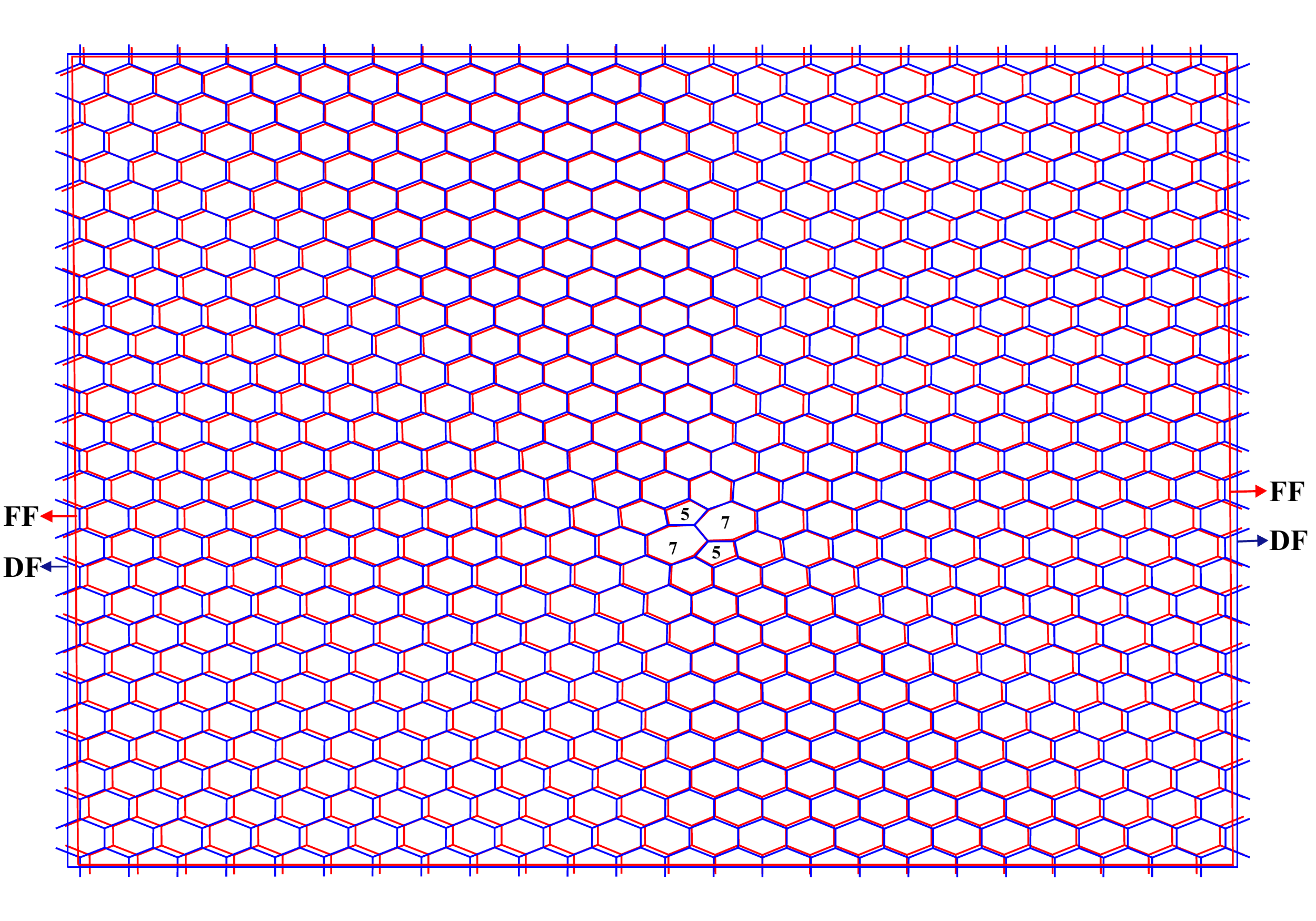}
\caption{Top view of a buckled 1344-atom graphene sample with a single SW defect, minimized with DF (blue) and FF (red) boundary conditions.}
\label{figure_S1}
\end{center}
\end{figure}

\begin{figure}
\begin{center}
\includegraphics[width=1.0\textwidth, height=5cm ]{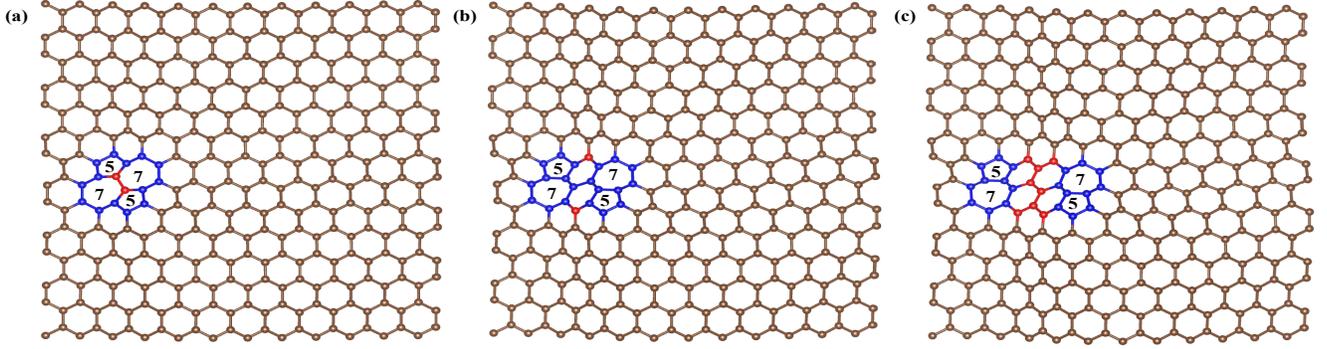}
\caption{Separation of the dislocations (pentagon-hexagon pairs)
by the introduction of $\Delta$ hexagon rings in between them.
(A)-(C) Dislocation separation varies from $\Delta$= $0$ to $2$.}
\label{figure_S2}
\end{center}
\end{figure}

\begin{figure}
\begin{center}
\includegraphics[width=1.0\textwidth ]{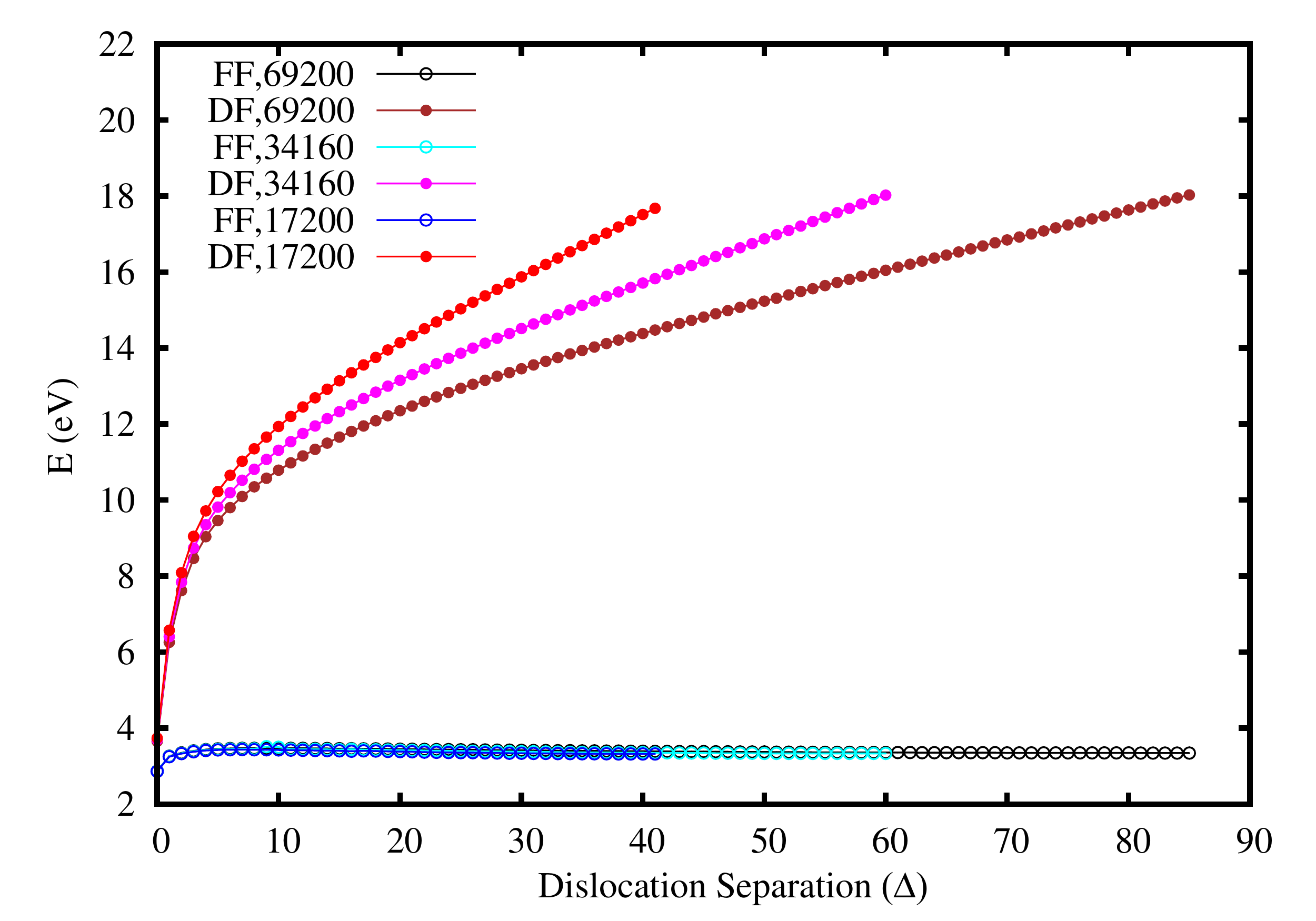}
\caption{The energy of a dislocation pair as a function
of the dislocation separation $\Delta$ for three different system
sizes, 17200, 34160 and 69200 carbon atoms, at two different boundary
conditions - FF and DF. Note that the formation energy difference between
FF and DF boudaries is more than 10 eV for the largest sample at large
dislocation separations.}
\label{figure_S3}
\end{center}
\end{figure}

$\bullet$\emph{\bf Computational details.}

To calculate the formation energies of defects in graphene,
we use a recently developed semiempirical potential given by [S1]
\begin{align}\label{potential}
E&=\frac{3}{16}\frac{\alpha}{d^2}\sum_{i,j}(r_{ij}^2-d^2)^2
  +\frac{3}{8}\beta d^2\sum_{j,i,k}\left(\theta_{jik}-\frac{2\pi}{3}\right)^2
  +\gamma\sum_{i,jkl}r_{i,jkl}^2.
\end{align}
Here, $r_{ij}$ is the distance between two bonded atoms, $\theta_{jik}$
is the angle between the two bonds connecting atom $i$ to atoms $j$
and $k$, and $r_{i,jkl}$ is the distance between atom $i$ and the
plane through the three atoms $j$, $k$ and $l$ bonded to atom $i$.
The parameters in the potential (\ref{potential}) are obtained by
fitting to density-functional theory (DFT) calculations: $d=1.420$
\AA~is the ideal bond length for graphene, $\alpha=26.060$~eV/\AA$^{2}$
is the bond-stretching constant and fitted to reproduce the bulk modulus,
$\beta=5.511$~eV/\AA$^{2}$ is the bond-shearing constant and fitted to
reproduce the shear modulus, and $\gamma=0.517$~eV/\AA$^{2}$ describes
the stability of the graphene sheet against buckling.

To prepare the graphene samples, first a supercell with periodicity vectors
$\vec{L}_x$ and $\vec{L}_y$ is created, in which $N$ carbon atoms are
placed according to the crystalline graphene structure. The defects are then introduced
in this crystalline sample, after which the atomic positions
are relaxed, i.e. the energy is minimized with the effective potential (\ref{potential}).
In the case of deformation free (DF) boundary conditions, the periodicity
vectors are kept fixed, while in the case of force free (FF) boundaries,
these vectors are allowed to adjust (their lengths as well as the angle
between them) in order to minimize the total energy. The structural
differences between DF and FF boundaries are shown in Fig.\ 4 for a
1344-atom sample with a single Stone-Wales (SW) defect.  Allowing the periodicity
vectors to relax lowers the energy. Therefore, the formation energy with
FF boundaries will always be lower than with DF boundaries. Moreover,
with increasing sample size, the difference between the two sets of
periodicity vectors vanishes, and the difference in formation energies
decreases to a nonzero offset.
\\

$\bullet$\emph{\bf Dislocation potential.} 

An SW defect can be considered as
a dislocation dipole [S2], in which a single dislocation is a
pentagon-heptagon pair. Two dislocations can be separated by introducing
hexagonal rings in between them as shown in Fig.\ 5. We calculated the
energy as a function of dislocation separation, measured as the number
$\Delta$ of introduced hexagonal rings for two different boundaries
(FF and DF) in the samples of three different sizes (17200, 34160 and
69200 atoms). Results are shown in Fig.\ 6. At large separation, the
energy difference between FF and DF boundaries reaches values of 10 eV or
more. This difference in energy shows that boundaries play crucial role
in the energetics of defects in graphene samples of the sizes used here.
\\

{\bf References:}

[S1] S.\ K.\ Jain, G.\ T.\ Barkema, N. Mousseau, C.-M. Fang, and M.\ A. van Huis,  J. Phys. Chem. C {\bf 119}, 9646 (2015).

[S2] A.\ Stone, D.\ Wales,  Chem. Phys. Lett. {\bf 128}, 501 (1986).

\end{document}